\documentclass[preprint,superscriptaddress,longbibliography,
nofootinbib,amsmath,amssymb,aps]{revtex4-1}

\usepackage{graphicx}
\usepackage{xcolor} 
\usepackage{subfigure}

\usepackage{amsmath,amssymb}
\usepackage{amsfonts}
\usepackage{natbib}
\usepackage{amssymb}
\usepackage{booktabs}
\usepackage{mathtools}
\usepackage[utf8]{inputenc}

\usepackage{tabularray}
\usepackage{tikz}

\newcommand{\sn}[1]{%
\begin{tikzpicture}[#1, scale = 1.5, line width = 0.4mm]%
\draw (0,0) -- (2ex,0);%
\draw[rounded corners=1.3ex] 
(2ex,1.3ex) -- (2ex,2ex) --  (4ex,2ex);%
\end{tikzpicture}}

\newcommand{\hopf}[1]{%
\begin{tikzpicture}[#1,scale = 1.5, line width = 0.4mm]%
\draw (0,0) -- (2ex,0);%
\draw[rounded corners=2.5ex] 
(2ex,0) -- (2ex,2ex) -- (4ex,2ex);%
\end{tikzpicture}}

\newcommand{\discont}[1]{%
\begin{tikzpicture}[#1,scale = 1.5, line width = 0.4mm]%
\draw (0,0) -- (2ex,0)  (2ex,1.5ex) --  (4ex,1.5ex);%
\end{tikzpicture}}

\begin{document}

\title{Using the basin entropy to explore bifurcations} 
\author{Alexandre~Wagemakers}
\email{Corresponding author: alexandre.wagemakers@urjc.es}
\affiliation{Nonlinear Dynamics, Chaos and Complex Systems Group, Departamento de  F\'isica, Universidad Rey Juan Carlos\\ Tulip\'an s/n, 28933 M\'ostoles, Madrid, Spain}

\author{Alvar~Daza}
 \affiliation{Nonlinear Dynamics, Chaos and Complex Systems Group, Departamento de  F\'isica, Universidad Rey Juan Carlos\\ Tulip\'an s/n, 28933 M\'ostoles, Madrid, Spain}

\author{Miguel A.F. Sanju\'{a}n}
\affiliation{Nonlinear Dynamics, Chaos and Complex Systems Group, Departamento de  F\'isica, Universidad Rey Juan Carlos\\ Tulip\'an s/n, 28933 M\'ostoles, Madrid, Spain}
\affiliation{Department of Applied Informatics, Kaunas University of Technology, Studentu 50-415, Kaunas LT-51368, Lithuania}
\date{\today}

\begin{abstract}
Bifurcation theory is the usual analytic approach to study the parameter space of a dynamical system. Despite the great power of prediction of these techniques, fundamental limitations appear during the study of a given problem. Nonlinear dynamical systems often hide their secrets and the ultimate resource is the numerical simulations of the equations. This paper presents a method to explore bifurcations by using the basin entropy. This measure of the unpredictability can detect transformations of phase space structures as a parameter evolves. We present several examples where the bifurcations in the parameter space have a quantitative effect on the basin entropy. Moreover, some transformations, such as the basin boundary metamorphoses, can be identified with the basin entropy but are not reflected in the bifurcation diagram. The correct interpretation of the basin entropy plotted as a parameter extends the numerical exploration of dynamical systems. 
\end{abstract}

\maketitle

\section{\label{sec:Introduction}Introduction}

At the heart of bifurcation theory along with the study of equilibria and their stability~\cite{wiggins2003introduction}, a typical question arises when we face the analysis of a certain dynamical system: ``What happens when we change a given parameter of the dynamical system?''. The powerful results obtained from bifurcation analysis allow us to understand the behavior of the system in parameter space. Nevertheless, using only a mathematical analysis is often limited and cannot always account for all the complex behaviors of nonlinear dynamical systems. Clearly, we need to use numerical tools to collect information about the system.

A few years ago, the new concept of basin entropy~\cite{daza2016basin,daza2018basin} was introduced, as a global measure of the unpredictability in the phase space of a given dynamical system. Numerically, it is expressed as a single value between $0$ and $\log(N_A)$, where $N_A$ is the number of asymptotic states in the region of phase space considered. The value $0$ means absolute predictability, that is, all the initial conditions end in the same final state with complete certainty. On the other end of the scale, the value $\log(N_A)$ indicates a completely fractalized phase space known as riddled or intermingled basin~\cite{daza2022classifying}, that is, the pinnacle of unpredictability. This simple idea of quantifying the unpredictability of a dynamical system through the entropy of its basins has brought a flourishing number of applications, both theoretical \cite{puy2021test, daza2022classifying} and practical \cite{halekotte2021transient}. A recent perspective article  glossing this same idea appears in \cite{dazaunprec2023}

The analysis of structural changes in the parameter space is the subject of bifurcation theory. Changes in the system parameters can affect the life of the usual inhabitants of the phase space, such as fixed points, invariant manifolds, attractors, and so on. These structures can appear and disappear, collide or change their stability. Needless to say, some of these transformations can affect the basins of attraction and therefore the basin entropy of the phase space. This is precisely our goal here: to explore the relation between the evolution of the basin entropy when a parameter is modified and certain types of bifurcations

The comparison between the classical bifurcation diagram as a function of a single parameter and the entropy of the basin calculated as a function of the same parameter provides information about the structure of the basins and also about the nature of the bifurcations. Although it cannot be considered as a rigorous analytical tool, when correctly interpreted, the basin entropy also reflects the occurrence of bifurcations in phase space.

However, we must keep in mind that the entropy of the basin is blind to bifurcations that change a single attractor into another, for example, a period doubling cascade. Since in this case, there is only a single basin, and as a consequence, this change will not be noted in the basin entropy calculation.

Definitely, it is not surprising that the qualitative changes in the phase space have their counterpart in the basin entropy. However, we must keep in mind that the basin entropy is blind to the bifurcations that change a single attractor into another, for example a period-doubling cascade. Since in this case, there is only a single basin, and as a consequence, this change will not be noted in the basin entropy computation of the basin.            

In short, our main goal is to propose a novel use for the basin entropy as a complementary tool for bifurcation analysis when multistability is involved. The visual inspection of a diagram for one or two parameters clearly indicates where the bifurcations are located and its effects on the basins.  

The article is organized as follows. In Section II, the relevant transitions in the phase space and their relation to the basin entropy are studied in detail. In Section III, we present an example of application to the well-known Hénon map. Some comments on the methods and the code availability are described in Section IV. Finally, we provide some concluding remarks at the end.

\section{Bifurcations and boundary metamorphoses}

Bifurcation theory is one of the pillars of the study of dynamical systems. In particular, it focuses on identifying the critical parameter values at which qualitative changes occur, and characterizing the new dynamics that emerge after the bifurcation. This involves analyzing the stability and bifurcation of equilibrium points, periodic orbits, and other invariant sets of the system. First, we will mainly deal with local bifurcations, such as saddle-node bifurcations, transcritical bifurcations, and pitchfork bifurcations, that are often characterized by the appearance or disappearance of equilibrium points or periodic orbits, as well as changes in their stability properties. Additionally, we also analyze some important cases of global bifurcations in the phase space. 

Our purpose now is to explore connections between results obtained from the analysis using bifurcation theory with those obtained by computing the basin entropy. By definition, the basin entropy takes an average over the region of the phase space under study. This means that its value reflects both local and global changes that occur as a parameter evolves. We will classify these changes into two groups. The first one concerns the number of coexisting stable states. For example, an asymptotic state can appear, disappear or change its stability as a consequence of a bifurcation. The second group will concern the transformation of the basins themselves, such as a transition from a smooth to a fractal boundary. These modifications are called basin boundary metamorphoses~\cite{grebogi1987basin} or basin bifurcations~\cite{mira1994basin}. 

We will study some examples referred to these two previous groups mentioned earlier and the basin entropy. Unquestionably, we do not claim to be exhaustive in the proposed list of studied examples. As a result, we have selected the most important local and global bifurcations affecting the multistability of a system. 

\subsection{Saddle-node bifurcation} 

The saddle-node bifurcation is the appearance of a saddle and a stable node as a parameter changes. It brings a new steady state to the phase space and can actually change the value of the basin entropy. We will consider a very simple example of a dynamical system that undergoes a saddle-node bifurcation. The Hénon map can be considered as a paradigm of two-dimensional discrete dynamical systems and, in fact, contains most of the behaviors of interest for this study.  We will use the notation as in \cite{grebogi1987basin}. 
\begin{equation}\label{eq:henon}
\begin{aligned}
x_{n+1} &= A - x_n^2 - J y_n\\
y_{n+1} &= x_n
\end{aligned}
\end{equation} 

Some analytical conditions must be fulfilled to guarantee a local behavior of the map similar to the normal form of the saddle-node bifurcation \cite{wiggins2003introduction}. For the Hénon map in Eq.~(\ref{eq:henon}), it is possible to obtain an accurate set of parameter values where these bifurcations happen. Notice that the analytical conditions are also valid for the fixed points of the iterated map $f^n$, then periodic orbits of the map $f$ can also undergo saddle-node bifurcations.

\begin{figure}

\includegraphics[]{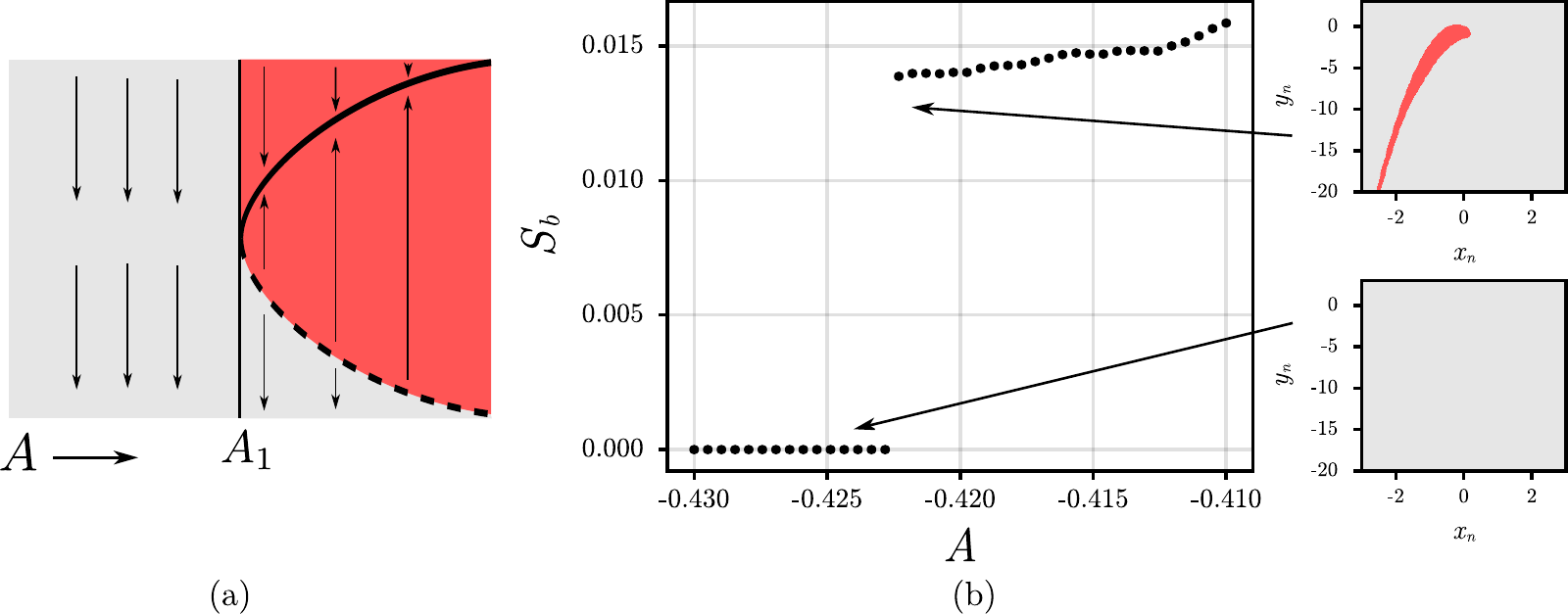}

\caption{{\bf Effects of the saddle-node bifurcation}. (a) Sketch of the basins and equilibrium states before and after the saddle-node bifurcation. A pair of saddle (dashed line) and node fixed points (solid line) appear at the bifurcation point. (b) Basin entropy for the Hénon map before and after the saddle-node bifurcation for the parameter values $A_1 = -0.422$ and $J=0.3$. Basins of attraction of the Hénon map are represented for $A^-_1 = -0.425$ (bottom panel) and $A^+_1 = -0.422$ (upper panel) to show the sudden appearance of the basin of the period-1 orbit.}
\label{fig1}
\end{figure}

\begin{figure}
\includegraphics[]{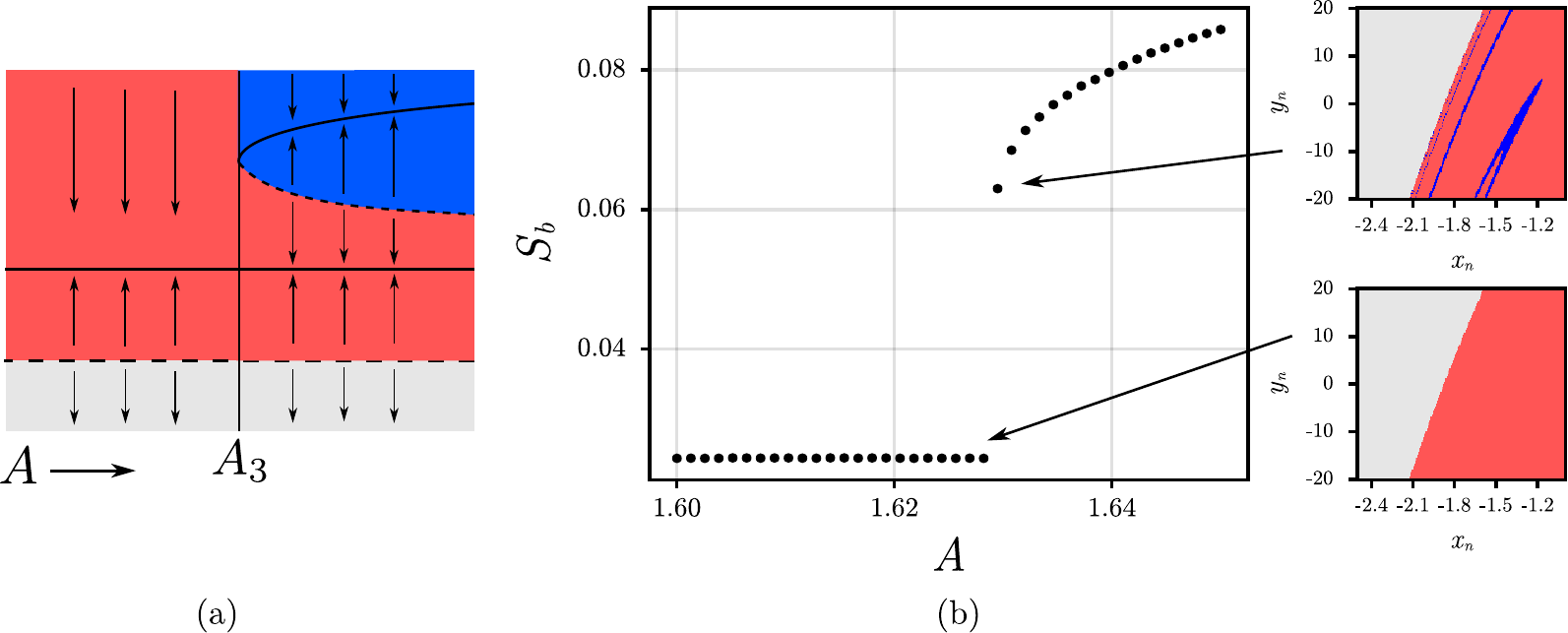}
\caption{{\bf Effects of the period-n saddle-node bifurcation}. (a) Sketch of the basins and equilibrium states before and after the saddle-node bifurcation. A pair of saddle (dashed line) and node fixed points (solid line) appear at the bifurcation point. (b) Basin entropy as a function of $A$ in the Hénon map for the period-3 saddle node bifurcation for the parameter values $A_3 \simeq 1.63$ and $J=0.05$. A detail of the basins are shown at $A_3^- = 1.6293745$ just before the bifurcation (bottom panel) and at $A_3^+ = 1.629375$ just after the bifurcation has occurred. The period-3 basin of attraction in blue appears just after the bifurcation.}
\label{fig2}
\end{figure}

We focus now on the basins of the phase space attractors before and after the saddle-node bifurcation has occurred. To make things clear, we refer to the situation {\it before} when there is no fixed point and {\it after} when the saddle and the stable fixed point coexist in phase space.   

The first saddle-node bifurcation for the system (\ref{eq:henon}) occurs at $A_1 = -(J+1)^2/4$ when a stable fixed point and a saddle appear. The basin entropy is discontinuous at this point and  will abruptly change through the passage of this bifurcation. We analyze the cause of this change:  
\begin{itemize}
\item $A < A_1$: Before the bifurcation, all initial conditions diverge and we consider that they belong to the same basin. In this case, the basin entropy is zero. 
\item $A > A_1$: The saddle appears on the boundary between the basin of the stable node and the divergent initial conditions. The stable manifold of the saddle forms the basin boundary.  At this point, the basin has a volume larger than zero and so does the basin entropy. However, depending on the region of the phase space the basin entropy can be arbitrarily small. 
\end{itemize} 
The basins of attraction and the basin entropy are shown in Fig. ~\ref{fig1}(b) for parameter values before and after the bifurcation. 

The analysis of the normal form $df/dx = a - x^2$ of the saddle-node bifurcation for ODEs \cite{wiggins2003introduction} leads to a similar analysis. The volume of the basin of the stable node grows discontinuously as shown in Fig.~\ref{fig1}(a). The boundary created by the saddle divides the phase space into two connected components. The same topological argument applies to the saddle-node bifurcation of the Hénon map, although the boundary has a very different shape.\\

Finally,  we turn our attention to saddle-node bifurcations on limit cycles and periodic orbits. We can track numerically the appearance of a period-3 orbit through a saddle-node bifurcation for the parameter values $A_3 = 1.62937(5)$ and $J = 0.05$. Although the basin of this new attractor is small in comparison to the main attractor, its appearance produces a small discontinuity in the value of the basin entropy. We will come back later on this period-3 orbit, since its disappearance occurs through a completely different process. The basins of attraction and the basin entropy near the saddle-node bifurcation is shown in Fig.~\ref{fig2}(b). The thin blue lines appearing in the red basin after the bifurcation corresponds to the basin of the new period-3 orbit.  

\subsection{Pitchfork bifurcation}

\begin{figure}
\includegraphics[]{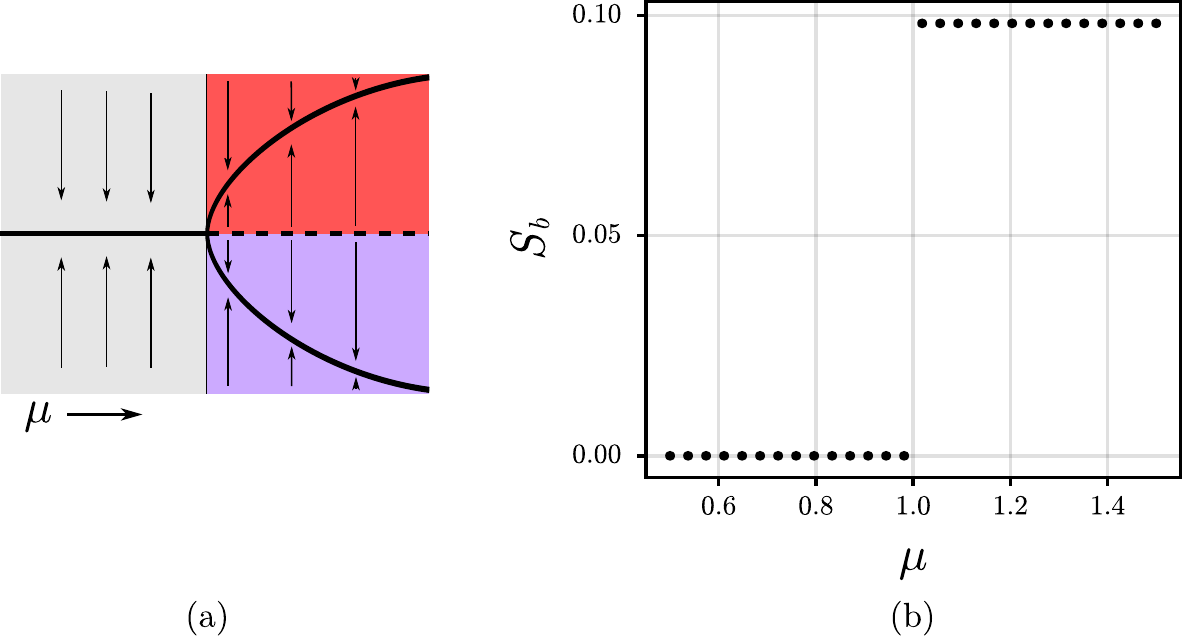}
\caption{{\bf Effects of the pitchfork bifurcation}. (a) Sketch of the basins and equilibrium states before and after a pitchfork bifurcation. The stable fixed point turns into a pair of stable fixed points separated by a saddle fixed point. The stable manifold of the saddle forms the boundary between the two basins. (b) Basin entropy for the cubic map (\ref{eq:cubic-map}) when the bifurcation occurs at $\mu = 1$. The transformation of the basins causes the discontinuous jump in the basin entropy at the bifurcation.} 
\label{fig3}
\end{figure}

The pitchfork bifurcation transforms a stable fixed point into a pair of stable fixed points plus a saddle fixed point. The phase space switches from one to two basins. An idealized phase space is represented in Fig.~\ref{fig3}(a) where the stable fixed points are pictured as solid thick lines and the dashed line holds for the unstable fixed point. After the bifurcation, the two basins are separated by the stable manifold of the saddle fixed point. 

The simplest map that fulfills the analytical conditions for the pitchfork bifurcation \cite{wiggins2003introduction} is the one-dimensional cubic map: 
\begin{equation} \label{eq:cubic-map}
x_{n+1} = \mu x_n - x_n^3
\end{equation} 

For $0 < \mu < 1$ there is a single stable fixed point at $x^* = 0$ and the basin entropy is zero for this parameter range. The bifurcation occurs at $\mu = 1$ and the two new stable fixed points appear for $\mu > 1$, see Fig.~\ref{fig3}(a). The basin entropy jumps to a positive value that will depend on the size of the observed phase space. Once again, we have a discontinuity in the basin entropy after the bifurcation point. If the observed region is symmetric around the origin, we have also a symmetry in the basins. The saddle separates the phase space in two parts just after the bifurcation.  

For continuous systems, there is an interesting case of a pitchfork bifurcation in the double-well Duffing oscillator. The stable equilibrium at the origin will turn into two stable fixed points separated by a boundary originated at the saddle fixed point. The basins are symmetric with an entangled smooth boundary. 

\subsection{Subcritical Hopf/Neimark-Sacker bifurcation}

\begin{figure}
\begin{center}	
\includegraphics[]{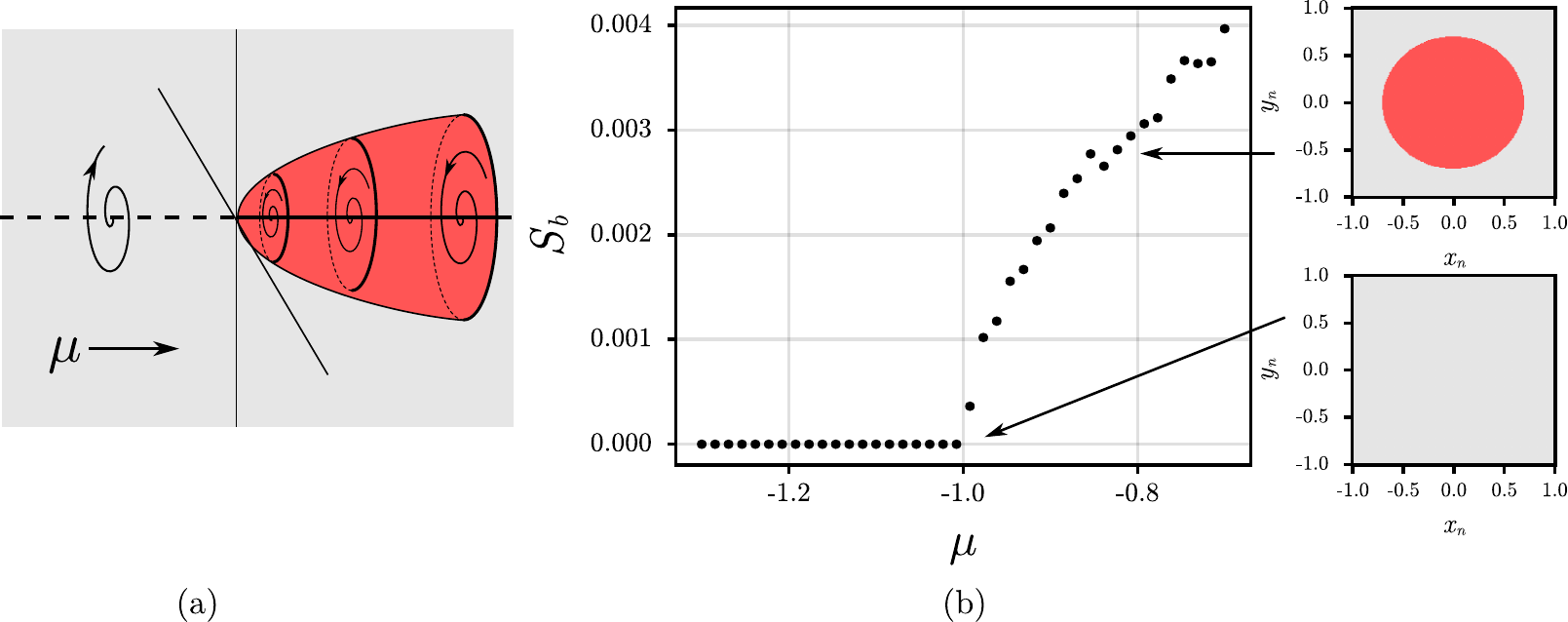}
\end{center}	
\caption{{\bf Effects of the sub-critical Hopf bifurcation}. (a) Sketch of the basins and equilibrium states before and after the Neimark-Sacker bifurcation. The unstable fixed point represented as a dashed line changes its stability at the bifurcation point. At the same time an unstable limit cycle appears around the stable fixed point. This cycle forms a boundary in the phase space. (b) Basin entropy for the map (\ref{eq:sub-hopf}) when the bifurcation occurs at $\mu = -1$. The unstable limit cycle appears at the bifurcation point in the phase space (the red dot). Its interior is the basin of attraction of the stable fixed point.} 
\label{fig4}
\end{figure}

When a subcritical Hopf bifurcation occurs, a stable fixed point coexists with an unstable cycle in the phase space until they collapse at the bifurcation point. It implies a change in the multistability of the system as the unstable cycle defines a boundary between the stable fixed point and the possible states outside the unstable limit cycle. In Fig.~\ref{fig4}(a), the unstable fixed point, painted as a dashed line, turns stable at the bifurcation parameter. At the same time, an unstable limit cycle appears around the stable fixed point. The size of the limit cycle is vanishing at the bifurcation point, but tends to grow as the bifurcation parameter increases. 

An interesting consequence of this merging of the two states is the fact that the basin volume of the fixed point shrinks until its disappearance at the bifurcation point. The basin entropy follows this tendency and decreases until reaching zero.

The following discrete map undergoes a Neimark-Sacker bifurcation which is somehow the equivalent of the Hopf bifurcation for a discrete map: 
\begin{equation}\label{eq:sub-hopf}
\begin{aligned}
x_{n+1} &= (\mu	- x_n^2 -y_n^2) x_n  - 0.1 y_n\\
y_{n+1} &=(\mu	- x_n^2 -y_n^2) y_n  + 0.1 x_n
\end{aligned}
\end{equation}  

For $\mu < -1$ all initial conditions diverge, and the subcritical Hopf bifuraction takes place at the value $\mu = -1$. The fixed point at the origin becomes stable and the unstable limit cycle begins to grow. As a consequence, the basin of attraction of the origin is exactly the surface enclosed by the limit cycle. Figure~\ref{fig4}(b) illustrates the evolution of the basin entropy as a function of the parameter $\mu$. 

This bifurcation is important in some neuronal dynamical systems where the bistability between a limit cycle and a stable fixed point allow bursting dynamics \cite{wage2006,ermentrout2010}. It has also been reported in discrete dynamical systems such as in the Bogdanov map \cite{arrowsmith1993}. 

\subsection{Boundary crisis and interior crisis}

\begin{figure}
\includegraphics[]{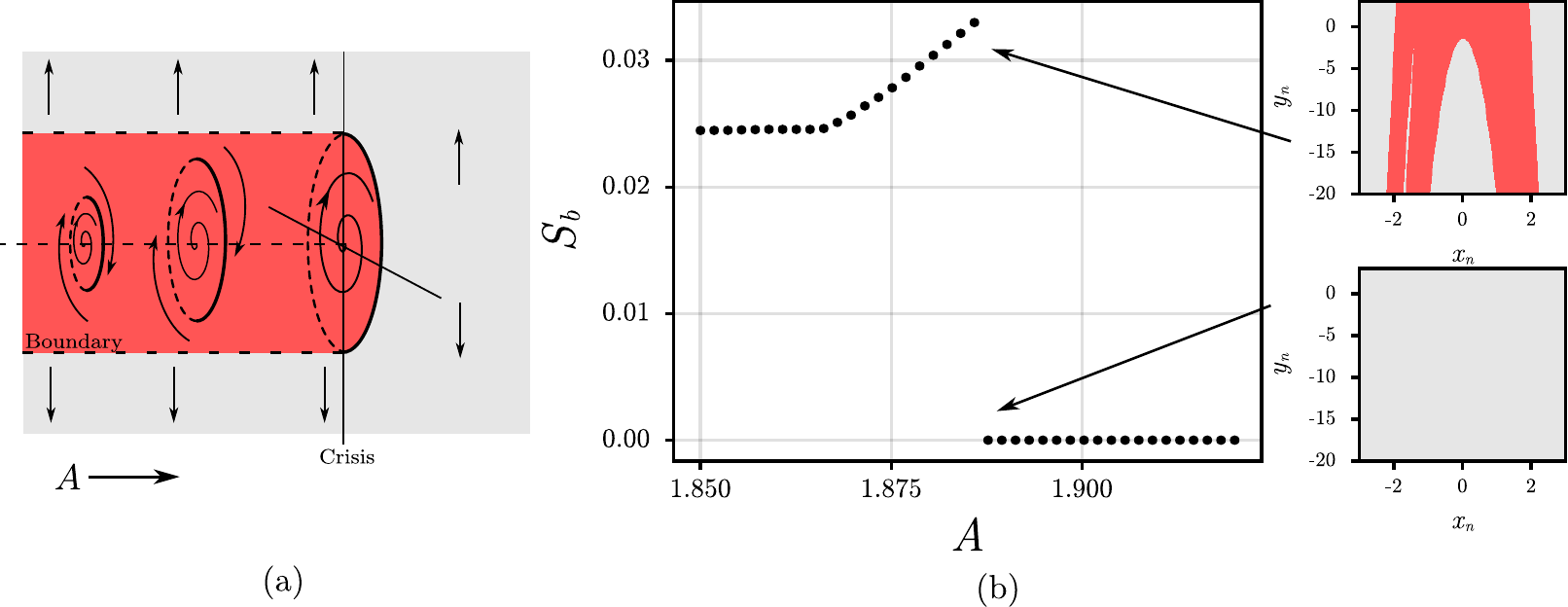}
\caption{{\bf Effects of the boundary crises}. (a) Sketch of the basins and equilibrium states before and after the boundary crises. The attractor grows inside the basin as the parameter $A$ increases. Eventually, it collides with the basin boundary when the crisis occurs. Its basin is destroyed and the initial conditions lead to another state. (b) Basin entropy for the map (\ref{eq:henon}) when the bifurcation occurs for the parameter values $A_1^c = 1.8874$ and $J=0.05$. This is the period-1 crisis boundary where the period-1 orbit touches the stable manifold of the period-1 saddle. Beyond this point all initial conditions diverge.} 
\label{fig5}
\end{figure}

Boundary crises \cite{grebogi1987basin,grebogi1983crises} occur when a stable dynamical state collides with an unstable state on the basin boundary as the parameter of the system increases, as sketched in Fig.~\ref{fig5}(a). It occurs for example in the Hénon map (\ref{eq:henon}) when periodic orbits or fixed points touch the unstable saddle on the boundary. As a result, a stable state is destroyed and it affects the multistability of the dynamical system.  Unlike the previous local bifurcations that study the local stability of the fixed point, this phenomenon is not easy to track analytically since it involves the size and position of an attractor in phase space.

Interior crises are a different type of crises, but lead to similar effects. The attractor collides with an unstable saddle inside the basin triggering the destruction of the attractor and its basin.  

This transition occurs several times in the Hénon map (\ref{eq:henon}) as the parameter $A$ evolves for $J=0.05$. At $A_1^c = 1.8874$, the chaotic attractor is tangent to the boundary and the crisis hits. The effect on the basin entropy is shown in Fig.~\ref{fig5}(b). 

The main result is a discontinuity of the basin entropy, since a whole basin disappears after the crisis, so the unpredictability should be reduced. Boundary crises have been reported in many continuous and discrete systems, since it is a very common event in parameter space.

\subsection{Homoclinic bifurcation}

\begin{figure}
\includegraphics[]{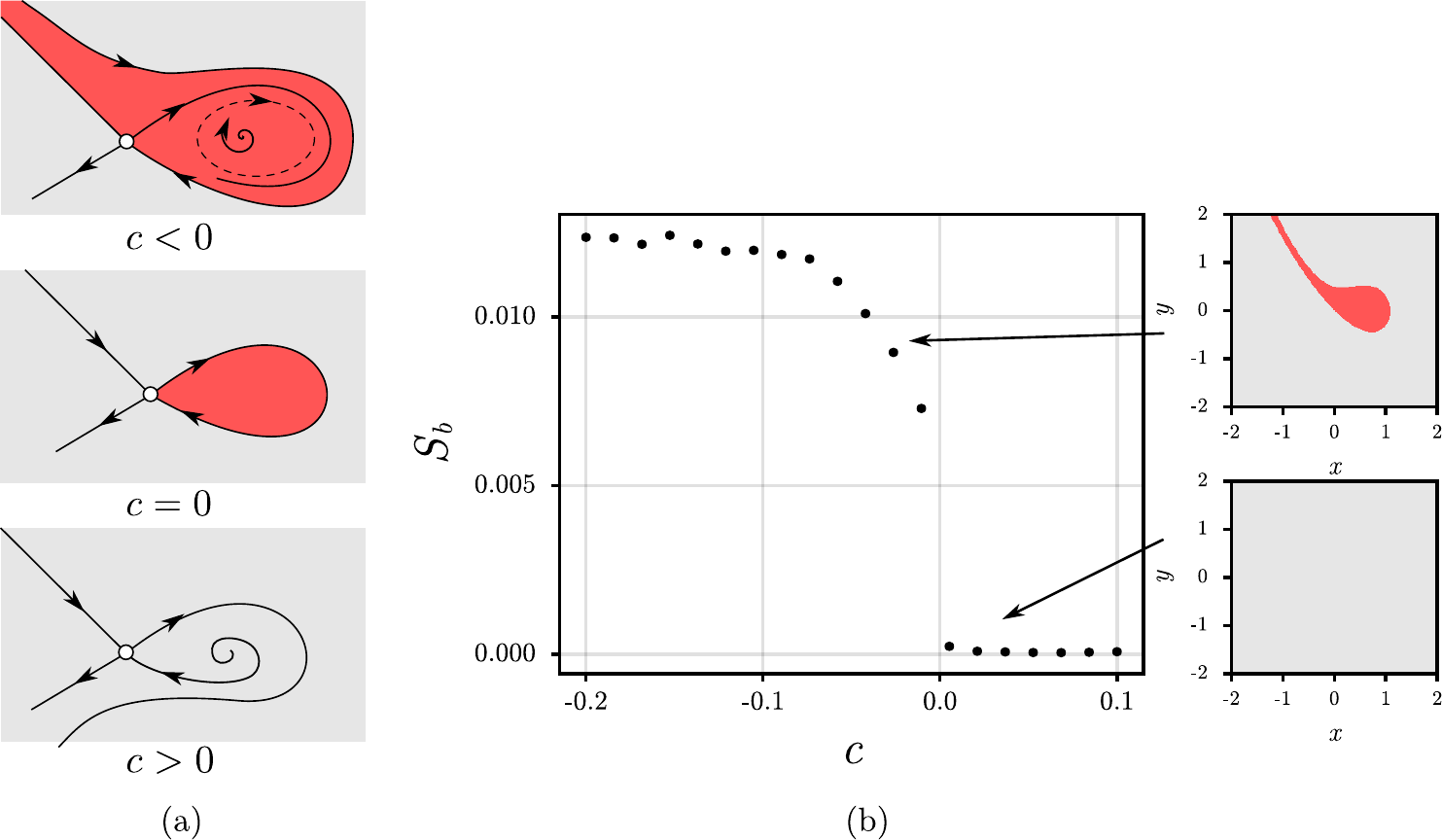}
\caption{{\bf Effects of the homoclinic bifurcation}. (a) Representation of the evolution of the homoclinic bifurcation. The limit cycle for $c<0$ is separated from the saddle through the unstable manifold that circle the saddle. At $c=0$ the two collide and the limit cycle disappears. For $c>0$ only the saddle remains. (b) Basin entropy for the map (\ref{eq:homoclinic}). There is a discontinuity at $c=0$ due to the disappearance of one of the basins. The panels show details of the basins of attraction before the homoclinic bifurcation at $c = -0.2$ and after at $c = 0.1$.} 
\label{fig6}
\end{figure}

When a limit cycle merges with an homoclinic loop of a saddle point, we have a homoclinic bifurcation in the phase space. When the saddle and the limit cycle coexist, we can consider that we have a bistable state where the stable manifold of the saddle separates the two basins. After the bifurcation, the limit cycle suddenly disappears and only the saddle remains. This process is illustrated in Fig.~\ref{fig6}(a) with a sketch of the phase space for three representative situations. The following ordinary differential equation proposed in \cite{hale2012dynamics} undergoes this bifurcation as the parameter $c$ increases: 
\begin{equation}\label{eq:homoclinic}
\begin{aligned}
    \dot x &= 2y\\
    \dot y &= 2x - 3x^2 -y(x^3 - x^2 + y^2 -c)
\end{aligned}
\end{equation}

For negative parameter values, $c <0$, we have the saddle and the limit cycle coexisting in the phase space. There is a smooth boundary between the two basins and the basin entropy is positive. After the homoclinic bifurcation, for positive parameter values $c>0$, the limit cycle no longer exists and only the saddle remains. There is only one basin and the basin entropy is zero.

In Fig.~\ref{fig5}(b), we represent the evolution of the basin entropy as the parameter $c$ increases as well as the basins before and after the bifurcation. Due to the disappearance of a stable state, the basin entropy is discontinuous at this point.

\subsection{Basin boundary metamorphoses}

We now turn our attention to the basin boundary metamorphoses. We focus on the effects on the morphology of the basins and their boundaries, when some structure embedded in the phase space evolves as a parameter changes. For this study, we consider a constant number of attractors in phase space. Although there are several metamorphoses described in the literature \cite{grebogi1987basin}, we will center our attention on the transitions due to the appearance of new fractal structures. 

\begin{figure}
\includegraphics[]{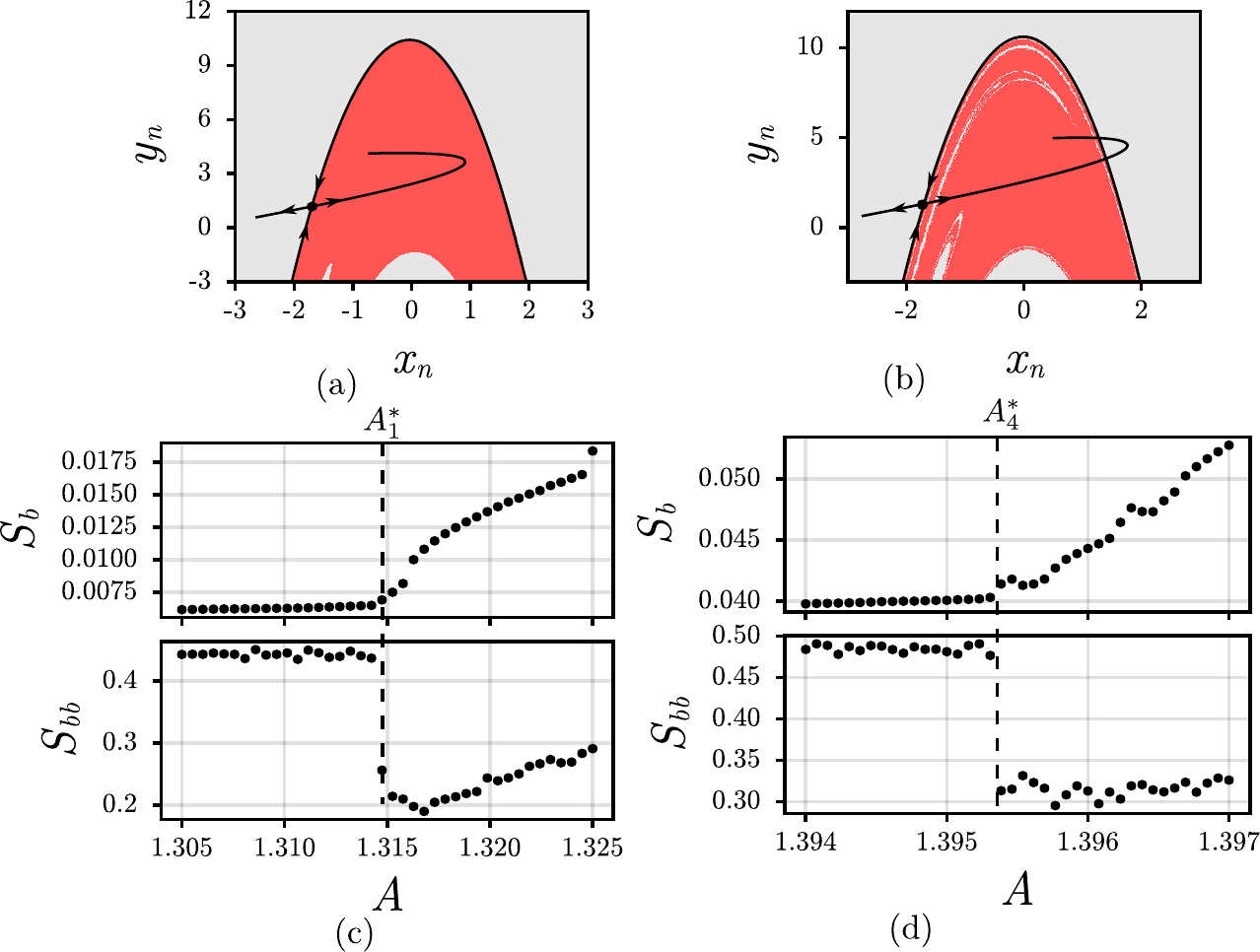}
\caption{{\bf Effects of the boundary metamorphoses}. (a) and (b) Sketch of the unstable and stable manifold of a period-1 saddle becoming tangent and its effect on the basin boundary. In (a) the boundary is smooth and the basin of the period-1 fixed point is depicted in red. The parameter values are $A = 1.305$, $J=0.3$. In (b) the unstable and stable manifold of the saddle point on the boundary have become tangent, causing the boundary to become suddenly fractal. The parameters values are $A = 1.36$, $J=0.3$. (c) Basin entropy for the map (\ref{eq:henon}) when the stable and unstable manifold of the period-1 saddle crosses at the transition $A_1^* = 1.315$. The basin switches from smooth to fractal and there is a sudden transition visible for $S_{bb}$. We have taken $J=0.3$ for this simulation. (d) Transition due to the period-4 saddle at $A_4^* = 1.395$. The boundary is already fractal, but the period-4 saddle homoclinic tangency increases the fractality of the boundary. This is a fractal-fractal basin boundary metamorphosis.} 
\label{fig7}
\end{figure}

A new fractal boundary can appear when the stable and unstable manifold of a periodic saddle becomes tangent. The boundary of the basins changes suddenly at this point as shown in Fig.~\ref{fig7}(a) and (b). The smooth-fractal metamorphosis, as the name suggests, turns a smooth boundary into a fractal. It can also occur when the boundary is already fractal, which means that additional fractal structures modify the current boundary. It is called fractal-fractal metamorphosis.  

For these transitions, it is helpful to use the basin boundary entropy $S_{bb}$ instead of the basin entropy. The computation of this quantity is equivalent to the basin entropy except that the average is only taken on the boxes lying on the boundary. It will be essential to grasp the details of the transformation on the boundary due to these homoclinic connections. 

We reproduce here the example of the Hénon map (\ref{eq:henon}) which has a transition of this nature for the parameter values $A_1^* = 1.315$, $J = 0.3$. The basin entropy and the boundary basin entropy are shown for a range of parameters around this value. The stable and unstable manifold of the period-1 saddle on the boundary become tangent (see Fig.~\ref{fig7}(a) and (b)), and this change is noticeable in two ways: 
\begin{itemize}
\item The basin entropy changes its tendency at this point. As $A$ increases the curve bends upward.  
\item The boundary basin entropy decreases brutally. However, at least numerically, the change seems continuous.
\end{itemize}

Other basin metamorphoses occur when period-n saddles on the boundary perform a homoclinic intersection. For the fractal-fractal basin boundary metamorphosis, the fractal basin suddenly changes its structure due to this new addition of filaments in the previous basin. For example at $A_4^* = 1.3965$, the stable and unstable period-4 saddle also becomes tangent. This change is more subtle and only leaves a small deviation in the basin entropy at this value as shown in Fig.~\ref{fig7}(d). Still the boundary basin entropy has a sharp drop at $A^* _4$.  

In \cite{shrimali2008nature}, the authors study these transitions with the uncertainty exponent of the basins \cite{grebogi_final_1983}. These changes clearly appear when the exponents are computed as a function of the chosen parameter. We claim that the basin entropy can achieve the same purpose but it is sensitive to other changes in the basins as well. For example, the uncertainty exponent is not affected to changes in the number of attractors.\\

A similar phenomenon to homoclinic tangency occurs for noninvertible maps when special structures called critical curves collide with the basin boundary. These transitions are called basin bifurcations in \cite{mira1994basin} and the basins can be transformed in different ways: 
\begin{itemize}
\item Connected basins $\longleftrightarrow$ Disconnected basins.
\item Simply connected $\longleftrightarrow$ Multiply connected basins. 
\item Smooth boundary $\longleftrightarrow$ Fractal boundary. 
\end{itemize}

In \cite{mira1994basin,shrimali2005basin}, the authors present a study of a noninvertible map presenting this kind of bifurcations. The basin entropy will change when the system goes through one of these transitions. We pick only a single example of transition in the quadratic map proposed in \cite{mira1994basin}: the smooth-fractal basin bifurcation. Consider the quadratic map: 
\begin{equation}\label{eq:quadratic}
\begin{aligned}
x_{n+1} &= a x_n + y_n \\
y_{n+1} &= x_n^2 +b
\end{aligned}
\end{equation}

Critical curves are the image of a special set $LC_{-1}$, which is the set of points where the Jacobian of the map vanishes. For a map $M: \mathbb{R}^2 \to \mathbb{R}^2$, the set $LC_n$ is the image $LC_n  = M(LC_{n-1})$. The tangency of these curves with the boundary causes the transformation of the basins called the basin bifurcation. 

\begin{figure}
\includegraphics[]{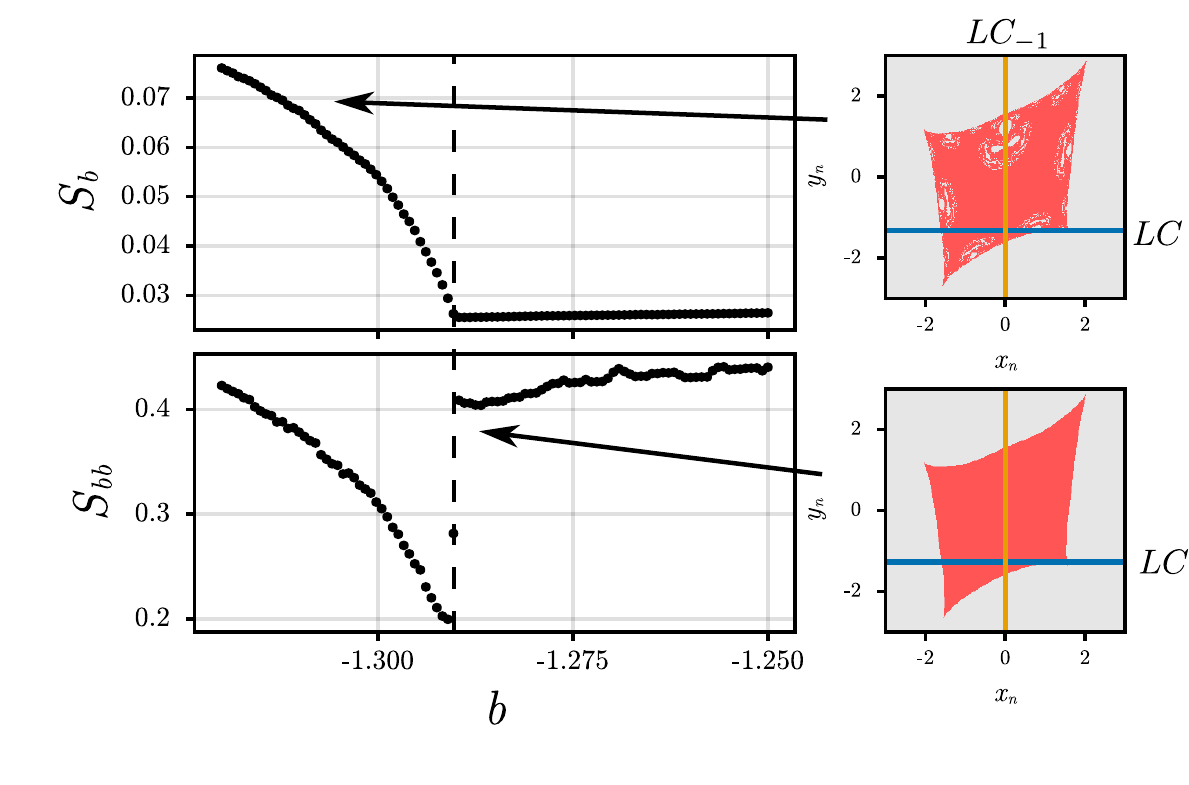}
\caption{{\bf Effects of the basin bifurcation}. Basin entropy and basin boundary entropy of the quadratic map (\ref{eq:quadratic}) as a function of the parameter $b$ for $a= -0.43$. The sharp transition around $b=-1.29$ for the basin entropy and the boundary basin entropy corresponds to the contact of a critical curve of the map with the basin boundary. The basins of attraction before (lower right panel) and after (upper right panel) show the effect of the tangency of the $LC_{-1}$ curve with the boundary.}
\label{fig8}
\end{figure}

In Fig.~\ref{fig8} the transition from smooth to fractal is depicted as a function of the parameter $b$. The transition is clearly visible around $b=-1.29$, where the critical curve $LC$ is tangent and then intersects with the basin boundary creating a fractal pattern. $S_b$ and $S_{bb}$ are plotted as a function of $b$. The sudden transition at $b = -1.29$ for the basin boundary entropy is a sign of a smooth-fractal transition. Moreover, the basin entropy changes its tendency and quickly increases in the same fashion as the boundary metamorphoses shown in Fig.~\ref{fig7}~(c).

\section{Bifurcation diagram of the Hénon map}

\begin{figure}
\includegraphics[width=0.85\textwidth]{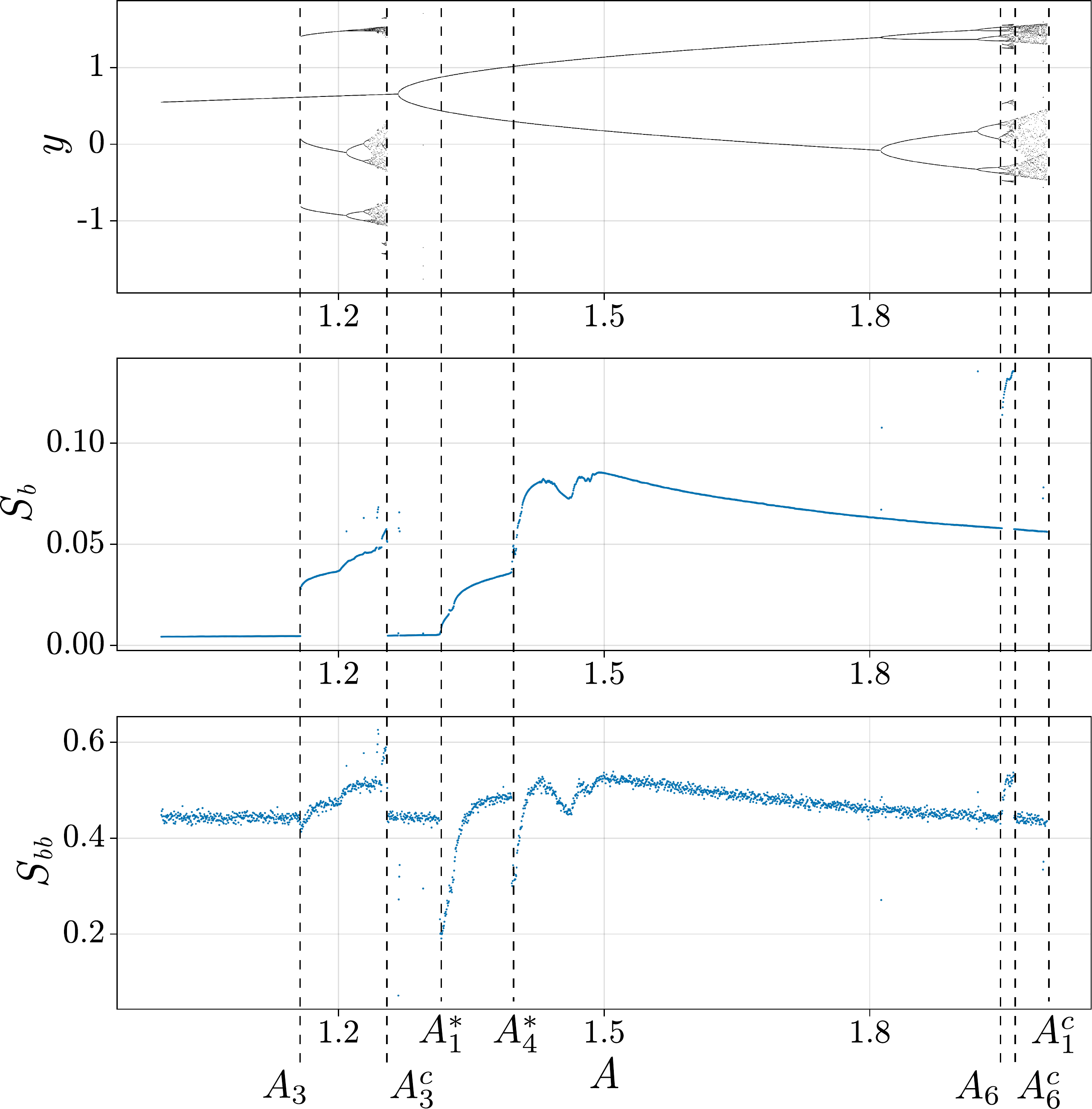}
\caption{Bifurcation diagram of the Hénon map $(x_n, y_n) \to (A -x_n^2 + Jy_n, x_n)$ for the parameter $J=0.3$. The events marked with dashed lines represent the appearance of period-n orbits $A_n$, destruction of attractors through boundary crises $A_n^c$, and the homoclinic tangency $A_n^*$. The basin entropy and the boundary basin entropy can help to identify the events occurring in phase space. The attractors and the entropies have been computed on the grid $(-3; 3)\times (-3;12)$ with a resolution of $5000\times5000$ regularly spaced initial conditions for $2000$ values of the parameter $A$.}
\label{fig9}
\end{figure}

The previous observations are applied here to the the analysis of the bifurcation diagram with the help of the basin entropy and boundary basin entropy. Figure~\ref{fig9} represents the bifurcation for the Hénon map (\ref{eq:henon}) as a function of the parameter $A$ for $J=0.3$. We have noted on the diagram several events that have been described earlier: 
\begin{itemize}
\item $A_n$: appearance of a period-n orbit through a saddle-node bifurcation. 
\item $A_n^c$: destruction of an attractor through a boundary crisis of a period-n saddle. 
\item $A_n^*$: homoclinic tangency of the period-n saddle. These are the smooth-fractal and the fractal-fractal transitions.
\end{itemize}
For this map, there is no basin bifurcation possible since the map has a constant Jacobian by construction. 

The graph of the bifurcation diagram reveals important dynamical transformations of the attractors such as the period-doubling cascade. But some of the events are not detected in the bifurcation diagram such as the homoclinic tangency of the saddle that transforms the boundary. We want to bring the attention on the behavior of the basins near a saddle-node bifurcation. It has been shown \cite{shrimali2008nature} that the size of the basin after the bifurcation increases as a power-law of the bifurcation parameter. This swelling of the basin is noticeable in the basin entropy. Just after the bifurcation $A_3$ and $A_6$, there is a jump in the value of the basin entropy as predicted, and then $S_b$ smoothly increases with the parameter $A$. This fact helps to locate such bifurcations and improves the readability of the diagram. 

The basin entropy cannot replace the continuation softwares for the bifurcation in parameter space. However, the computation of the basins can bring to light some attractors that would have been undetected otherwise. Definitely, it is computationally more expensive, however, as a result we have a global picture of the phase space. We summarize in Tab. ~\ref{table1} the behavior of the basin entropy for the different cases we have enumerated in this article. 

\begin{table}
\begin{tblr}{
  hlines,
  rows = {ht=0.8cm},
  columns = {halign=c},
  } 
Phase space event & $S_b$ & $S_{bb}$ \\ 
\hline
Saddle node bifurcation & \sn{} &  \sn{} \\
Pitchfork bifurcation & \discont{} & \discont{}  \\
Subcritical Hopf bifurcation & \hopf{} & \hopf{}  \\
Boundary crises & \discont{} & \discont{}  \\
Homoclinic bifurcation & \discont{} & \discont{}  \\
Fractal-fractal metamorphosis & \hopf{} & \discont{}  \\
Smooth-fractal metamorphosis & \hopf{} & \discont{}  \\
Basin bifurcation & \hopf{} & \discont{}  \\
\end{tblr}
\caption{\label{table1} Summary of the effects of the different events in the phase space on the basin entropy and basin boundary entropy. The symbols represents visual cues for the transitions that the entropy experiences through the bifurcation.}
\end{table}

\section{\label{methods}Methods and code availability}
All the simulations have been performed using the Julia programming language, computed with an automatic algorithm described in \cite{datseris2022effortless}. The program first computes the basins on the chosen grid and the basin entropy and the basin boundary entropy are estimated. The resolution and the grid size used for the basins are specified in the figure captions. The integrator chosen is a 9th order Verner method with relative tolerance set to $1\cdot 10^{-9}$. The code necessary to generate the figures and to compute the basins is available at \cite{coderef}.

\section{Conclusion}
The main goal of this research work is to show how the study of dynamical systems from the point of view of their basins of attraction can unveil a trove of different bifurcations in parameter space. In other words, the classical bifurcation diagram and continuation software do not reveal all the information on the evolution of the basins and the basin entropy can detect these changes. 

Furthermore, the basin entropy helps to identify bifurcations in the phase space and helps to classify their type. While it is not a fully reliable method to classify systematically the bifurcations, it contributes to bring some possible candidates to the front. Along with other measures, it can be considered as a qualitative method to apply when the basins of attraction have been computed. In any case, it is a helpful additional tool available to researchers in dynamical systems to help investigating the phase space.  

\begin{acknowledgments}
This work has been supported by the Spanish State Research Agency (AEI) and the European Regional Development Fund (ERDF, EU) under Project No.~PID2019-105554GB-I00 (MCIN/AEI/10.13039/501100011033)
\end{acknowledgments}


\begin{thebibliography}{99}

\bibitem{wiggins2003introduction}
Wiggins S. Introduction to applied nonlinear dynamical systems and chaos,  volume~2. Springer; 2003.

\bibitem{daza2016basin}
Daza A, Wagemakers A, Georgeot B, Gu{\'e}ry-Odelin D, Sanju{\'a}n MAF. Basin entropy: a new tool to analyze uncertainty in dynamical systems. Sci. Rep.. 2016; 6:1--10.

\bibitem{daza2018basin}
Daza A, Wagemakers A, Georgeot B, Gu{\'e}ry-Odelin D, Sanju{\'a}n MAF. Basin entropy, a measure of final state unpredictability and its application to the chaotic scattering of cold atoms. In: Chaotic, Fractional, and Complex Dynamics: New Insights and Perspectives, p. 9--34. Springer; 2018.

\bibitem{daza2022classifying}
Daza A, Wagemakers A, Sanju{\'a}n MAF. Classifying basins of attraction using the basin entropy. Chaos, Solitons \& Fractals. 2022; 159:112112.

\bibitem{puy2021test}
Puy A, Daza A, Wagemakers A, Sanju{\'a}n MAF. A test for fractal boundaries based on the basin entropy. Commun Nonlinear Sci Numer Simul. 2021; 95:105588.

\bibitem{halekotte2021transient}
Halekotte L, Vanselow A, Feudel U. Transient chaos enforces uncertainty in the british power grid. Journal of Physics: Complexity. 2021; 2:035015. 

\bibitem{dazaunprec2023}
Daza A, Wagemakers A, Sanju{\'a}n MAF. Unpredictability and basin entropy. Europhysics Letters. 2023; 141(4):43001.

\bibitem{grebogi1987basin}
Grebogi C, Ott E, Yorke JA. Basin boundary metamorphoses: changes in accessible boundary orbits. Nuclear Physics B-Proceedings Supplements. 1987; 2:281--300.

\bibitem{mira1994basin}
Mira C, Fournier-Prunaret D, Gardini L, Kawakami H, Cathala JC. Basin bifurcations of two-dimensional noninvertible maps: fractalization of basins. International Journal of Bifurcation and Chaos. 1994; 4(02):343--381.

\bibitem{wage2006}
Wagemakers A, Sanju{\'a}n MAF, Casado JM, Aihara K. Building electronic bursters with the Morris--Lecar neuron model. International Journal of Bifurcation and Chaos. 2006; 16(12):3617--3630.

\bibitem{ermentrout2010}
Ermentrout  B, Terman DH. Mathematical foundations of neuroscience, volume~35. Springer; 2010.

\bibitem{arrowsmith1993}
Arrowsmith DK, Cartwright JHE, Lansbury AN, Place CM. The Bogdanov map: Bifurcations, mode locking, and chaos in a dissipative system. International Journal of Bifurcation and Chaos. 1993; 3(04):803--842.

\bibitem{grebogi1983crises}
Grebogi C, Ott E, Yorke JA. Crises, sudden changes in chaotic attractors, and transient chaos. Physica D: Nonlinear Phenomena. 1983; 7(1-3):181--200.

\bibitem{hale2012dynamics}
Hale JK, Ko{\c{c}}ak H. Dynamics and bifurcations, volume~3. Springer; 2012.

\bibitem{shrimali2008nature}
Shrimali MD, Prasad A, Ramaswamy R, Feudel U. The nature of attractor basins in multistable systems. International Journal of Bifurcation and Chaos. 2008; 18(06):1675--1688.

\bibitem{grebogi_final_1983}
Grebogi C, McDonald SW, Ott E, Yorke JA. Final state sensitivity: An obstruction to predictability. Phys. Lett. A. 1982; 99:415--418.

\bibitem{shrimali2005basin}
Shrimali MD, Prasad A, Ramaswamy R, Feudel U. Basin bifurcations in quasiperiodically forced coupled systems. Physical Review E. 2005; 72(3):036215.

\bibitem{datseris2022effortless}
Datseris  G, Wagemakers A. Effortless estimation of basins of attraction. Chaos: An Interdisciplinary Journal of Nonlinear Science. 2022; 32(2):023104.

\bibitem{coderef}
{Code available at Github repository}.
 \url{https://github.com/awage/BifurcationBasinEntropy}.
\end{thebibliography}

\end{document}